\documentclass[11pt,a4paper,twoside]{article}
\usepackage{amsmath}
\usepackage{amsfonts}
\usepackage{changes}
\usepackage{hyperref}
\numberwithin{equation}{section}

\begin{document}

\noindent

{\bf
{\Large Conserved chiral currents on the boundary of 3D Maxwell theory \\ 

}
} 

\vspace{.5cm}
\hrule

\vspace{1cm}

\noindent

{\large\bf{Nicola Maggiore\footnote{\tt nicola.maggiore@ge.infn.it }
\\[1cm]}}

\setcounter{footnote}{0}

\noindent
{{}Dipartimento di Fisica, Universit\`a di Genova,\\
via Dodecaneso 33, I-16146, Genova, Italy\\
and\\
{} I.N.F.N. - Sezione di Genova\\
\vspace{1cm}

\noindent
{\tt Abstract~:}

In this paper the 3D Maxwell theory with single-sided planar boundary is studied. As a consequence of the existence, on the boundary, of two Ward identities, we find two chiral conserved edge currents satisfying a Ka\c{c}-Moody algebra with central charge equal to the inverse of the Maxwell coupling constant. We show that the boundary degrees of freedom are two 2D scalar chiral bosons whose chiralities depend on the parameters of the bulk Maxwell theory. In particular, the edge chiral bosons may have opposite chiralities, in close analogy with the ``spinon'' and ``holon'' currents characterizing the 3D topological insulators.

\newpage

\section{Introduction}

One of the first studies on the consequences of the presence of a boundary in field theory was done by Casimir in \cite{Casimir:1948dh}, where the pressure between two parallel conducting plates was computed, as an example of modification of the vacuum state of the electromagnetic field. Later, in \cite{Symanzik:1981wd}, Symanzik faced the general problem of introducing and studying the effect of a boundary in a Quantum Field Theory (QFT), starting from the definition itself of boundary. Symanzik's basic idea was that a boundary separates the space into two regions. The field theoretical translation of this defining property concerns the propagators of the theory, which are constrained to vanish between points lying on different sides of the boundary (separability constraint). This approach has also the advantage of solving the ambiguity related to the boundary conditions which must be satisfied by the fields of the theory. According Symanzik's procedure, indeed, the boundary conditions are fixed by the separability constraint on the propagators which, in turn, are  derived from the Ward identities of the theory. Hence, the boundary conditions are ultimately determined by the symmetries of the theory. Symanzik's approach has been successfully applied in situations where the boundary plays a role in QFT, and its effects can be measured \cite{Blasi:1992mm}. Particularly interesting is the case of Topological QFTs (TQFTs), which do not possess any local observable, the only cohomologically nontrivial objects being globally defined, like for instance Wilson loops, or surfaces, or geometrical knots \cite{Witten:1988ze}. Local physical observables of TQFTs are confined on the boundary, if present. For instance, introducing a boundary in 3D Chern-Simons (CS) theory serves to classify all rational conformal field theories \cite{Moore:1989yh}, and chiral conserved currents are derived on its edge, which form a Ka\c{c}-Moody (KM) algebra with central charge related to the CS coupling constant \cite{Blasi:1990pf,Emery:1991tf,Maggiore:2017vjf}. This nontrivial boundary structure has a physical interpretation in the Fractional Quantum Hall Effect \cite{Wen:1990se,Blasi:2008gt}. Similarly, on the boundary of topological BF theory \cite{Birmingham:1991ty}, which can be defined in any spacetime dimensions, KM algebras satisfied by conserved chiral currents are found \cite{Amoretti:2012kb,Amoretti:2013nv}, thus justifying the claim according to which topological BF theories are the effective field theories of topological insulators \cite{Hasan:2010xy,Cho:2010rk}. It makes difference if the boundary is double or single-sided. Of this latter type, for instance, are the theories involved in the AdS/CFT correspondence \cite{Maldacena:1997re}: a gravity theory in the $(D+1)$-dimensional bulk holographically reduces to a gauge theory on its (single-sided)  $D$-dimensional boundary. This mechanism is also known as ``gauge-gravity duality'', and it has important applications also beyond its original string theory framework \cite{Hartnoll:2009sz,McGreevy:2009xe,Amoretti:2017xto}. By extension, this correspondence can also be referred to flat bulk theories, not involving gravity \cite{2013ass,Amoretti:2014kba}, with a single-sided boundary \cite{Amoretti:2014iza}. In these cases, Symanzik's method can be superseded by the introduction of a theta step function in the action, without need of imposing the separability constraint on the propagators, which might be a difficult task. The results may be different, as is the case, for instance, of topologically massive Maxwell-CS (MCS) theory \cite{Deser:1981wh}: while the physics on the boundary turns out to be independent from the Maxwell term in the case of double-sided boundary studied {\it \`a la} Symanzik \cite{Blasi:2010gw}, for single-sided boundary the central charge of the KM algebra and the chiral velocities of the conserved chiral edge currents depend on the Maxwell coupling constant \cite{Maggiore:2018bxr}. While  the CS case can be recovered from MCS theory as the zero limit of the Maxwell coupling constant, the zero limit of the CS coupling constant diverges, and the 3D Maxwell theory with single-sided boundary must be treated separately, and this is the subject of this paper, which is organized as follows. In Section 2 the 3D Maxwell action with planar single-sided boundary is written, together with the gauge fixing term corresponding to the axial gauge, the external term where external sources are coupled to the fields of the theory,  and the most general boundary term . In Section 3 the boundary conditions on the gauge fields are derived from their equations of motion, and the Ward identities are written. We find that two Ward identities hold, which is peculiar to the Maxwell theory alone, since only one Ward identity exists when the Maxwell action is coupled to a CS term \cite{Maggiore:2018bxr}. From the two Ward identities, the existence of two conserved currents is deduced, which satisfy a KM algebra with central extension equal to the inverse of the Maxwell coupling constant, in surprising analogy with the 3D topological BF theory with boundary \cite{Maggiore:1992ki,Blasi:2011pf}. In Section 4 we show that the degrees of freedom of the 2D boundary are two scalar fields, which are constrained by the boundary conditions on the bulk gauge fields found previously. We study two particular solutions with physical relevance, and we find that the constraints can be solved by chiral scalar fields with opposite chiralities, which renders the analogy with the 3D topological BF theory complete. Moreover, since this latter model is proposed as the effective field theory for 3D topological insulators \cite{Cho:2010rk}, this result suggests the unforeseen and interesting possibility  of an alternative field theoretical description of 3D topological insulators, by means of the Maxwell field theory. In Section 5 we summarize and further comment our results.

\section{The action with boundary}

The Maxwell action defined on a closed, flat euclidean half-space is given by
\begin{equation}
S_{M}(A) = -\frac{\kappa}{4} \int d^3x\ \theta(x_2) \ F_{\mu\nu} F_{\mu\nu} \ ,
\label{2.1}\end{equation}
where $F_{\mu\nu}(x)$ is the electromagnetic field strength ($F_{\mu\nu}=\partial_\mu A_\nu-\partial_\nu A_\mu$), $\theta(x_2)$ is the step function which introduces in the theory the single-sided planar boundary
\begin{equation}
x_2=0,
\label{2.2}\end{equation}
and $\kappa$ is a real positive constant which could be normalized to one by a redefinition of the gauge field $A_\mu(x)$, which nevertheless we prefer to keep generically positive, in order to be able to identify at any time the role of the Maxwell term in what follows. In 3D, the canonical mass dimension of the gauge field is $[A]=1/2$. Our notations are as follows: Greek letters run over the 3D euclidean spacetime ($\mu=0,1,2$), while Latin letters cover the 2D boundary $x_2=0$ ($i=0,1$). Moreover, the 3D coordinates are denoted by $x=(x_0,x_1,x_2)$, while the plane $x_2=0$ is spanned by $X=(x_0,x_1)$. Accordingly, the Maxwell action \eqref{2.1} reads
\begin{eqnarray}
S_M(A) &=&
-\frac{\kappa}{4} \int d^3x\ \theta(x_2)\  (F_{ij} F_{ij} + 2F_{2i} F_{2i})
\nonumber\\
&=& 
-\frac{\kappa}{2} \int d^3x\theta(x_2)
\left[
(\partial_0A_1)^2
+ (\partial_1A_0)^2
+ (\partial_0A_2)^2
+ (\partial_2A_0)^2
+ (\partial_2A_1)^2
\right.\nonumber\\
&&\left.
-2 \partial_0A_1\partial_1A_0
-2 \partial_0A_2\partial_2A_0
- 2 \partial_1A_2\partial_2A_1
+ (\partial_1A_2)^2
\right].
\label{2.3}
\end{eqnarray}
A convenient choice for the gauge fixing is the axial one
\begin{equation}
A_2=0,
\label{2.4}\end{equation}
which is implemented by the gauge fixing term
\begin{equation}
S_{gf}=\int d^3x\ \theta(x_2)bA_2 \ ,
\label{2.5}\end{equation}
where $b(x)$ is a Lagrangian multiplier, a.k.a. Nakanishi-Lautrup multiplier \cite{Nakanishi:1966zz,Lautrup:1967zz}. As usual, external fields are coupled to the gauge fields:
\begin{equation}
S_{\gamma}=\int d^3x\ \left[\theta(x_2)\gamma_i A_i +\delta(x_2)\gamma_{2i}(\partial_2A_i)\right]\ .
\label{2.6}\end{equation}
Notice that in $S_\gamma$ the external sources $\gamma_i(x)$ are coupled only to the gauge fields $A_i(x)$, and not to $A_2(x)$, which is gauged away by the axial gauge choice \eqref{2.4}. In addition, sources $\gamma_{2i}(x)$ must be coupled also to $\left.\partial_2A_i(x)\right|_{x_2=0}$ since, on the boundary $x_2=0$, the fields $\left.A_i(x)\right|_{x_2=0}$ and their $\partial_2$-derivatives $\left.\partial_2A_i(x)\right|_{x_2=0}$ are independent one from each other. 
We stress that the fields $A_i(x)$ and their $\partial_2$-derivatives, are not independent dynamical variables {\it in the half 3D space (described by $x=(x_0,x_1,x_2)$},  but the physics {\it on the 2D plane $X=(x_0,x_1,x_2=0)$} depends on the type of boundary conditions the fields $A_i(x)$ undergo (a physical example is  the role which the  boundary conditions play in the field-theoretic description of the Casimir effect \cite{Asorey:2006pr}), and is described by the 2D generating functional ${\cal Z}[J(X)]$, where, in order to be able to treat independently Dirichlet and Neumann boundary conditions for the bulk fields, 
external sources must be coupled to both $\left.A_i(x)\right|_{x_2=0}$ and $\left.(\partial_2A_i(x))\right|_{x_2=0}$, so that the boundary conditions are effectively treated as dynamical fields \cite{Karabali:2015epa}.

To highlight this remark, from now on we adopt the following shorthand notation
\begin{equation}
\partial_2A_i(x) \equiv A_{i,2}(x)\ .
\label{2.7}\end{equation}
Correspondingly, the fields and their $\partial_2$-derivatives may undergo independent boundary conditions on $x_2=0$: of the Dirichlet, Neumann or Robin type. Finally, on $x_2=0$ a boundary action should be defined, whose most general form is
\begin{eqnarray}
S_{bd}&=&\int d^3x\ \delta(x_2)\left(
a_1A_0^2
+a_2A_0A_1
+a_3A_1^2
+a_4\partial_0A_0A_1
+a_5\partial_1A_0A_1 \right.
\nonumber \\
&&\left. +a_6A_{0,2}A_0
+a_7A_{0,2}A_1
+a_8A_{1,2}A_0
+a_9A_{1,2}A_1
\right)\ ,
\label{2.8}\end{eqnarray}
where $a_i,\ i=(1,\ldots,9)$ are constant parameters and the power counting prescription has been taken into account, together with the observation that, as already remarked, on the boundary $x_2=0$ $A_i(x)$ and $A_{2i}(x)$ must be considered as independent fields, and therefore $\partial_2$-integration by parts is not allowed. 
The 3D Maxwell theory is peculiar, because of power counting. In fact, the canonical mass dimensions of the 3D Maxwell gauge field $A_i(x)$ is one half, while for Chern-Simons theory, where only one spacetime derivatives appears, the dimension of the gauge field is one. As a consequence of the power counting constraint, in Chern-Simons theory the boundary term $S_{bd}$ cannot depend on $\partial_2 A_i$. This is not the case for 3D Maxwell theory, where, on its 2D boundary, $\partial_2 A_i$ may couple with $A_i$ forming a composite operator with mass dimension two. This seemengly technical remark has a deep physical consequence, as we will show in what follows. 

Notice that the boundary term $S_{bd}$ in its general form \eqref{2.8} breaks $O(2)$ invariance on the boundary. The 2D invariance under rotations can be recovered for the particular choice of the parameters $a_i$
\begin{equation}
a_1=a_3\ ;\ a_6=a_9\ ;\ a_7+a_8=0\ ;\ a_2=a_4=a_5=0.
\label{3.3}\end{equation}
The parameters $a_i$ appearing in $S_{bd}$ \eqref{2.8} will be determined to describe physically relevant situations on the 2D boundary of 3D Maxwell theory. We therefore prefer to maintain a general ground without imposing a symmetry which might turn out to be too restrictive. Indeed, it can easily be seen that asking that the boundary term is invariant under 2D rotations, by fixing the parameters $a_i$ to the choice \eqref{3.3}, would prevent from constructing relevant 2D dynamics. This is in close analogy to what happens in massive gravity \cite{Hinterbichler:2011tt}, where a Lorentz covariant mass term is not the most general choice \cite{Rubakov:2004eb,Libanov:2005vu}, so that alternative to the Fierz-Pauli mass terms \cite{Fierz:1939ix} are possible \cite{Blasi:2015lrg,Blasi:2017pkk}.

Summarizing, the total classical action for Maxwell theory built on a closed  flat half-space is given by
\begin{equation}
S_{tot}=S_M(A)+S_{gf}+S_\gamma+S_{bd}.
\label{2.9}\end{equation}

\section{Boundary conditions, Ward identities, conserved currents and Ka\c{c}-Moody algebra}

From the action $S_{tot}$ \eqref{2.9}, the field equations of motion are derived
\begin{eqnarray}
\frac{\delta S_{tot}}{\delta A_0} &=& 
\theta(x_2)(\kappa \partial_1 F_{10}+\kappa\partial_2F_{20}+\gamma_0)  
+\delta(x_2)(\kappa F_{20}+2a_1A_0+a_2A_1
\nonumber \\
&&
+a_6A_{0,2}
-a_4\partial_0A_1-a_5\partial_1A_1
+a_8A_{1,2}) \label{3.1}\\
\frac{\delta S_{tot}}{\delta A_1} &=& 
\theta(x_2)(\kappa \partial_0 F_{01}+\kappa\partial_2F_{21}+\gamma_1)  
+\delta(x_2)(\kappa F_{21}+a_2A_0+2a_3A_1
\nonumber \\
&&
+a_9A_{1,2}
+a_4\partial_0A_0+a_5\partial_1A_0
+a_7A_{0,2}). \label{3.2}
\end{eqnarray}
The boundary conditions on $x_2=0$ are obtained from the field equations \eqref{3.1} and \eqref{3.2} by integrating $\int_0^\epsilon dx_2$, letting $\epsilon\rightarrow 0$ and then going on-shell, $i.e.$ at vanishing external sources:
\begin{eqnarray}
2a_1A_0+a_2A_1+(a_6+\kappa)A_{0,2}
-a_4\partial_0A_1-a_5\partial_1A_1
+a_8A_{1,2} &=& 0\label{3.4}\\
a_2A_0+2a_3A_1+(a_9+\kappa)A_{1,2} 
+a_4\partial_0A_0+a_5\partial_1A_0 + a_7A_{0,2} &=&0\label{3.5}.
\end{eqnarray}
From the equations of motion \eqref{3.1} and \eqref{3.2} the following integrated Ward identity is also derived
\begin{equation}
\int_0^\infty dx_2 \partial_i\gamma_i = \left. \kappa\partial_iA_{i,2}\right|_{x_2=0}.
\label{3.6}\end{equation}
Having in mind the previous remarks, we compute the equations of motion of the fields $A_{i,2}(x)$:
\begin{eqnarray}
\frac{\delta S_{tot}}{\delta A_{0,2}} &=&
\theta(x_2)(\kappa\partial_0A_2-\kappa A_{0,2})
+\delta(x_2)(\gamma_{20}+a_6A_0+a_7A_1) \label{3.7}\\
\frac{\delta S_{tot}}{\delta A_{1,2}} &=&
\theta(x_2)(\kappa\partial_1A_2-\kappa A_{1,2})
+\delta(x_2)(\gamma_{21}+a_8A_0+a_9A_1)\label{3.8},
\end{eqnarray}
from which the additional on-shell boundary conditions follow
\begin{eqnarray}
a_6A_0+a_7A_1 &=& 0\label{3.9}\\
a_8A_0+a_9A_1 &=& 0\label{3.10},
\end{eqnarray}
and the local Ward identity on the boundary $x_2=0$
\begin{equation}
\left.\partial_i\gamma_{2i}\right|_{x_2=0} = -\left. \kappa\partial_iA_i\right|_{x_2=0}.
\label{3.11}\end{equation}

The additional local Ward identity \eqref{3.11} will play a crucial role, and it is important to stress that its existence  is peculiar to the Maxwell theory: it does not hold, indeed, when a Maxwell term is coupled, for instance, to CS theory \cite{Blasi:2010gw,Maggiore:2018bxr}. 
It is interesting to remark the presence, on the boundary $x_2=0$, of a local Ward identity. It is known, in fact, that the axial gauge does not completely fix the gauge symmetry, and that the residual gauge symmetry is described by a local Ward identity \cite{Bassetto:1991ue}. Moreover, local Ward identities in 2D ($i.e.$ on the boundary of a 3D theory) determine the form of the commutators of fields and currents, which translate into the 2D conformal algebra \cite{Fradkin:1997df}. Here something similar happens, as explained below.

The Ward identities \eqref{3.6} and \eqref{3.11}, taken on-shell, $i.e.$ at vanishing external sources $\gamma_i(x)=\gamma_{2i}(x)=0$, give rise to two conserved currents, which live on the planar boundary $x_2=0$:
\begin{eqnarray}
\left.\partial_iA_{i,2}(x)\right|_{x_2=0} &=& 0 \label{3.12} \\
\left.\partial_iA_i(x)\right|_{x_2=0} &=& 0. \label{3.13} 
\end{eqnarray}
The Ward identities \eqref{3.6} and \eqref{3.11} lead also to a 2D algebra of the conserved currents identified by \eqref{3.12} and \eqref{3.13}. In fact, deriving the Ward identity \eqref{3.6} with respect to $\gamma_k(x')$,
going at vanishing external sources and defining the ``time'' ordered  product with respect to the coordinate $x_0$, we get
\begin{eqnarray}
\partial_k\delta^{(2)}(X-X') &=& 
\kappa \partial_i\left\langle A_{i,2}(X))A_k(X')\right\rangle\nonumber\\
&=&\kappa\delta(x_0-x_0')\left[A_{0,2}(X),A_k(X')\right]\nonumber\\
&&
+\kappa\left\langle\partial_iA_{i,2}(X)A_k(X')\right\rangle \ .
\label{3.14}\end{eqnarray}
The last term on the right hand side of \eqref{3.14} is a contact term (c.t.) which vanishes using the conservation relation \eqref{3.12}, so that we get the 2D commutation relation
\begin{equation}
\left[A_{0,2}(X),A_1(X')\right]\delta(x_0-x_0')  = \frac{1}{\kappa}\partial_1\delta^{(2)}(X-X').
\label{3.15}\end{equation}
Similarly, deriving the Ward identity \eqref{3.6} with respect to $\gamma_{2k}(x')$ we get
\begin{eqnarray}
0&=& 
\kappa\partial_i\left\langle A_{i,2}(X)A_{k,2}(X')\right\rangle\nonumber\\
&=&\kappa\delta(x_0-x_0')\left[A_{0,2}(X),A_{k,2}(X')\right]+\mbox{c.t.}
\label{3.16}\end{eqnarray}
Deriving then the Ward identity \eqref{3.11} with respect to $\gamma_k(x')$ and $\gamma_{2k}(x')$ we get, respectively
\begin{eqnarray}
0&=& 
-\kappa \partial_i\left\langle A_i(X)A_k(X')\right\rangle\nonumber\\
&=&-\kappa\delta(x_0-x_0')\left[A_0(X),A_k(X')\right]+\mbox{c.t.}
\label{3.17}\end{eqnarray}
and
\begin{eqnarray}
\partial_k\delta^{(2)}(X-X') &=& 
-\kappa \partial_i\left\langle A_i(X)A_{k,2}(X')\right\rangle\nonumber\\
&=&-\kappa\delta(x_0-x_0')\left[A_0(X),A_{k,2}(X')\right]+\mbox{c.t.} 
\label{3.18}\end{eqnarray}
Summarizing, we find the following equal ``time'' $x_0=x'_0$ algebra formed by the conserved currents $A_{i,2}(X)$ \eqref{3.12} and $A_i(X)$ \eqref{3.13} on the boundary $x_2=0$ of the 3D Maxwell theory\footnote{The following properties of the delta function have been used: $\delta(-x)=\delta(x)$ and $\delta'(-x)=-\delta'(x)$.}
\begin{equation}
[A_1(X),A_{0,2}(X')] = \frac{1}{\kappa} \partial_1 (x_1-x'_1) 
\label{3.19}\end{equation}
\begin{equation}
[A_{0,2}(X),A_{k,2}(X')] = 0
\label{3.20}\end{equation}
\begin{equation}
[A_0(X),A_k(X')] = 0
\label{3.21}\end{equation}
\begin{equation}
[A_0(X),A_{1,2}(X')] = -\frac{1}{\kappa} \partial_1 (x_1-x'_1),
\label{3.22}\end{equation}
which can be written in a more compact way as
\begin{equation}
[A_i(X),A_{j,2}(X')] = -\frac{1}{\kappa}\epsilon_{ij}\partial_1\delta(x_1-x'_1),
\label{3.23}\end{equation}
taken at equal ``time'' $x_0=x_0'$.
The algebra \eqref{3.23} is a semidirect sum of KM algebras satisfied by the 2D conserved currents $A_i(X)$ and $A_{i,2}(X)$, with central extension $c=\frac{1}{\kappa}$, where $\kappa$ is the Maxwell coupling constant, in the mixed commutators. We  notice that this algebra coincides with the one which is found on the boundary of the 3D topological BF theory \cite{Maggiore:1992ki,Blasi:2011pf} , where the role of the second gauge field $B_\mu(x)$ is here played by $A_{\mu,2}(x)$. This is a crucial fact, which has important physical consequences for the boundary dynamics of the theory, the first of which is the interpretation of the conserved currents \eqref{3.13} and \eqref{3.12} as charge current  and spin current, respectively. These physical identifications will be discussed in the next Section.

\section{Boundary dynamics}

The current conservation relations \eqref{3.12} and \eqref{3.13} define two scalar fields on the 2D boundary $x_2=0$ of the 3D bulk Maxwell theory:
\begin{eqnarray}
\left. \partial_iA_i(x)\right|_{x_2=0} = 0 &\Rightarrow& A_i(X)=\epsilon_{ij}\partial_j\Phi(X) \label{4.1}\\
\left. \partial_iA_{i,2}(x)\right|_{x_2=0} = 0 &\Rightarrow& A_{i,2}(X)=\epsilon_{ij}\partial_j\Psi(X) \label{4.2}
\end{eqnarray} 
The scalar fields $\Phi(X)$ and $\Psi(X)$ are the 2D dynamical degrees of freedom which concern the physics on the boundary of the 3D Maxwell theory. The boundary conditions \eqref{3.4}, \eqref{3.5}, \eqref{3.9} and \eqref{3.10}, written in terms of the 2D scalar fields $\Phi(X)$ and $\Psi(X)$ are:
\begin{equation}
2a_1\partial_1\Phi-a_2\partial_0\Phi+(a_6+\kappa)\partial_1\Psi+\partial_0(a_4\partial_0\Phi+a_5\partial_1\Phi)-a_8\partial_0\Psi = 0 
\label{4.3}\end{equation}
\begin{equation}
a_2\partial_1\Phi-2a_3\partial_0\Phi-(a_9+\kappa)\partial_0\Psi+\partial_1(a_4\partial_0\Phi+a_5\partial_1\Phi)+a_7\partial_1\Psi = 0 
\label{4.4}\end{equation}
\begin{eqnarray}
a_6\partial_1\Phi-a_7\partial_0\Phi &=& 0 \label{4.5} \\
a_8\partial_1\Phi-a_9\partial_0\Phi &=& 0 \label{4.6}
\end{eqnarray}
Notice that the equations \eqref{4.5} and \eqref{4.6} describe a chiral scalar field $\Phi(X)$
\begin{equation}
\partial_0\Phi+v_\Phi\partial_1\Phi =0
\label{4.7}\end{equation}
which propagates in the $x_1$-direction with velocity $v_\Phi$. We give now two examples, tuned by the parameters $a_i$ appearing in $S_{bd}$ \eqref{2.8}, of the physics which occurs on the boundary of the 3D Maxwell theory. 

An interesting boundary situation is obtained taking
\begin{equation}
a_4=a_5=a_8=a_9=0.
\label{4.8}\end{equation}
The boundary conditions reduce to
\begin{eqnarray}
2a_1\partial_1\Phi-a_2\partial_0\Phi+(a_6+\kappa)\partial_1\Psi &=&  0 \label{4.9} \\
a_2\partial_1\Phi-2a_3\partial_0\Phi-\kappa\partial_0\Psi+a_7\partial_1\Psi &=& 0 \label{4.10} \\
a_6\partial_1\Phi-a_7\partial_0\Phi &=& 0. \label{4.11}
\end{eqnarray}
The condition \eqref{4.11} describes a chiral scalar field $\Phi(X)$ propagating in the $x_1$-direction with velocity
\begin{equation}
v_\Phi = -\frac{a_6}{a_7}.
\label{4.12}\end{equation}
Using \eqref{4.11}, the conditions \eqref{4.9} and \eqref{4.10} become
\begin{eqnarray}
(2a_1+a_2v_\Phi)\partial_1\Phi+(a_6+\kappa)\partial_1\Psi &=&  0 \label{4.13} \\
(a_2+2a_3v_\Phi)\partial_1\Phi-\kappa\partial_0\Psi+a_7\partial_1\Psi &=& 0. \label{4.14}
\end{eqnarray}
If, for instance, 
\begin{equation}
2a_1+a_2v_\phi=a_6+\kappa=a_2+2a_3v_\phi=0,
\label{4.15}\end{equation}
we are left with 
\begin{equation}
\partial_0\Psi-\frac{a_7}{\kappa}\partial_1\Psi =0,
\label{4.16}\end{equation}
which tells us that also the scalar $\Psi(X)$ is a chiral boson  with velocity
\begin{equation}
v_\psi=-\frac{a_7}{\kappa},
\label{4.17}\end{equation}
while, because of \eqref{4.11}, the velocity of the chiral boson $\Phi(X)$ is
\begin{equation}
v_\phi=\frac{\kappa}{a_7}.
\label{4.18}\end{equation}
This configuration describes two 2D chiral bosons $\Phi(X)$ and $\Psi(X)$ which propagate on the boundary $x_2=0$ of the 3D Maxwell theory with opposite and inverse velocities: 
\begin{equation}
v_\Phi v_\Psi = \left(-\frac{a_6}{a_7}\right)\left(-\frac{a_7}{\kappa}\right)=-1,
\label{4.19}\end{equation}
where we used \eqref{4.15}. The physical situation we are encountering here\footnote{We point out that the same physical result can be obtained with other choices of the parameters $a_i$.} reminds that of topological insulators \cite{Hasan:2010xy}, characterized by two chiral excitations moving on the edge of a surface with opposite velocities \cite{Cho:2010rk,Blasi:2011pf}. This allows the interpretation of the 2D conserved chiral currents \eqref{4.1} and \eqref{4.2} as charge current and spin current, respectively. The new fact here is that this peculiar feature of topological insulators is found on the boundary of 3D Maxwell theory, while the known fact so far is that this kind of behavior is recovered on the edge of 3D topological BF theory. Two quite different bulk theories (the topological BF theory and the non-topological Maxwell theory) on the boundary display the same physics (of topological insulators).

However, the dynamics which can be found on the boundary of 3D Maxwell theory is much richer, since the boundary conditions of the bulk theory may have more general solutions. For instance, let us consider again the four boundary conditions written in terms of the scalar fields $\Phi(X)$ and $\Psi(X)$: \eqref{4.3}, \eqref{4.4}, \eqref{4.5} and \eqref{4.6}. Like before, we chose $a_4=a_5=a_8=a_9=0$ and \eqref{4.5} describes a chiral boson $\Phi(X)$ with velocity $v_\Phi=-a_6/a_7$ \eqref{4.12}. We are then left with \eqref{4.13} and \eqref{4.14} as in the previous example. As before, let us take $a_2+2a_3v_\Phi=0$ \eqref{4.15}, so that \eqref{4.14} describes the chiral boson $\Psi(X)$ with chiral velocity $v_\Psi$ \eqref{4.17}. Eq. \eqref{4.13} represents a Robin boundary condition for the gauge field $A_0$:
\begin{equation}
\left.(2a_1+a_2v_\Phi)A_0+(a_6+\kappa)A_{0,2}\right|_{x_2=0}=0,
\label{4.20}\end{equation}
where we reintroduced the bulk gauge field $A_0$ through \eqref{4.1} and \eqref{4.2}. In other words, the set of boundary conditions can be solved by two chiral bosons provided that a Robin boundary condition on the gauge field $A_0(x)$ is given. A remark concerning the nature of the boundary degrees of freedom might be worth. In \cite{Amoretti:2013xya} it has been shown that the boundary conditions which are derived for the fields of topological BF theories in various spacetime dimensions fall into the class of duality conditions studied in \cite{Aratyn:1984jz,Aratyn:1983bg} which lead to a fermionization of the bosonic degrees of freedom involved. This means that the edge bosons for which a duality constraint holds, describe indeed fermionic degrees of freedom, which fits with the observation that the edge excitations of topological insulators are fermionic, despite the fact that the fields of the bulk effective field theory (3D BF theory) are bosonic. A similar fermionization mechanism might occurs for 3D Maxwell theory, where the role of the duality/boundary constraint studied in \cite{Amoretti:2013xya,Aratyn:1984jz,Aratyn:1983bg} is played by the Robin boundary condition \eqref{4.20}.
With respect to the previous case describing chiral bosons with opposite velocities, this time we have more choices, since
\begin{equation}
v_\Phi v_\Psi= \left(-\frac{a_6}{a_7}\right)\left(-\frac{a_7}{\kappa}\right)=\frac{a_6}{\kappa} 
\left\{
\begin{array}{ll}
>0 & \mbox{if}\ a_6>0 \\
<0 & \mbox{if}\ a_6 < 0
\end{array}
\right.
\label{4.21}\end{equation}
This solution therefore describes chiral bosons which may have opposite or concordant velocities, depending on the sign of one of the parameters appearing in the boundary term $S_{bd}$ \eqref{2.8}. This situation generalizes the one described in the previous example for topological insulators, where the conserved chiral currents travels in opposite directions on the edge.

\section{Conclusions}

In this paper we studied the 3D Maxwell theory with a single-sided planar boundary. Following standard methods of QFT, we found the most general boundary conditions compatible with the equations of motion of the gauge field $A_\mu(x)$, which depend on the parameters appearing in the boundary term of the total action. On the 2D boundary, the theory displays two Ward identities, which define two 2D conserved currents, obeying a KM algebra with central extension depending on the Maxwell coupling constant. The current conservation relations yield two 2D scalar fields, which are the dynamical variables on the 2D boundary. The boundary conditions on the gauge fields translate into constraints on the scalar fields, which can be solved by chiral bosons moving on the boundary of the 3D Maxwell theory with velocities which depend on the Maxwell coupling constant and on the parameters appearing in the boundary action, which may be tuned in order to have the chiral velocities either with the same or with the opposite sign. These features (conserved edge currents with opposite chiralities which obey a KM algebra with central charge equal to the inverse of the bulk coupling constant) coincide with those characterizing the boundary of 3D topological BF theory \cite{Maggiore:1992ki,Blasi:2011pf}, which is known to describe the physics of the topological insulators \cite{Cho:2010rk}. Maxwell and topological BF theory share the property of being gauge field theories respecting Time Reversal, which indeed plays an important role in the physics of topological insulators. Under any other respect, they are quite different theories, and having the same behavior on the boundary is an interesting result on its own. In the 3D topological BF theory, the two conserved edge chiral are interpreted as the currents of ``spinons'' and ``holons'' \cite{Cho:2010rk}. Pushing the analogy further, the role of the additional gauge field $B_\mu(x)$ is played here by the $\partial_2$-derivative of the gauge field $A_\mu(x)$, which, indeed, on the $x_2=0$ boundary must be considered as an independent field with respect to $A_\mu(x)$, obeying also independent boundary conditions (Dirichlet for $A_\mu(x)$, Neumann for $\partial_2A_\mu(x)$ and Robin for the mixed case). The analogy is impressive, and this new result deserves further studies before claiming that Maxwell theory with single-edged boundary represents an alternative effective theory for 3D topological insulators.


{\bf Acknowledgements}

The support of INFN Scientific Initiative SFT: ``Statistical Field Theory, Low-Dimensional Systems, Integrable Models and Applications'' is acknowledged.



\begin{thebibliography}{100}

\bibitem{Casimir:1948dh} 
  H.~B.~G.~Casimir,
  Indag.\ Math.\  {\bf 10}, 261 (1948)
  [Kon.\ Ned.\ Akad.\ Wetensch.\ Proc.\  {\bf 51}, 793 (1948)]
  [Front.\ Phys.\  {\bf 65}, 342 (1987)]
  [Kon.\ Ned.\ Akad.\ Wetensch.\ Proc.\  {\bf 100N3-4}, 61 (1997)].

\bibitem{Symanzik:1981wd} 
  K.~Symanzik,
  Nucl.\ Phys.\ B {\bf 190}, 1 (1981),
  doi:10.1016/0550-3213(81)90482-X.

\bibitem{Blasi:1992mm} 
  A.~Blasi, R.~Collina and J.~Sassarini,
  Int.\ J.\ Mod.\ Phys.\ A {\bf 9}, 1677 (1994).
  doi:10.1142/S0217751X94000728.

\bibitem{Witten:1988ze} 
  E.~Witten,
  Commun.\ Math.\ Phys.\  {\bf 117}, 353 (1988).
  doi:10.1007/BF01223371.

\bibitem{Moore:1989yh} 
  G.~W.~Moore and N.~Seiberg,
  Phys.\ Lett.\ B {\bf 220}, 422 (1989),
  doi:10.1016/0370-2693(89)90897-6.

\bibitem{Blasi:1990pf} 
  A.~Blasi and R.~Collina,
  Int.\ J.\ Mod.\ Phys.\ A {\bf 7}, 3083 (1992),
  doi:10.1142/S0217751X92001381.

\bibitem{Emery:1991tf} 
  S.~Emery and O.~Piguet,
  Helv.\ Phys.\ Acta {\bf 64}, 1256 (1991).

\bibitem{Maggiore:2017vjf} 
  N.~Maggiore,
  Int.\ J.\ Mod.\ Phys.\ A {\bf 33}, no. 02, 1850013 (2018)
  doi:10.1142/S0217751X18500136.

\bibitem{Wen:1990se} 
  X.~G.~Wen,
  Phys.\ Rev.\ B {\bf 41}, 12838 (1990),
  doi:10.1103/PhysRevB.41.12838.

\bibitem{Blasi:2008gt} 
  A.~Blasi, D.~Ferraro, N.~Maggiore, N.~Magnoli and M.~Sassetti,
  Annalen Phys.\  {\bf 17}, 885 (2008),
  doi:10.1002/andp.200810323.

\bibitem{Birmingham:1991ty} 
  D.~Birmingham, M.~Blau, M.~Rakowski and G.~Thompson,
  Phys.\ Rept.\  {\bf 209}, 129 (1991),
  doi:10.1016/0370-1573(91)90117-5.

\bibitem{Amoretti:2012kb} 
  A.~Amoretti, A.~Blasi, N.~Maggiore and N.~Magnoli,
  New J.\ Phys.\  {\bf 14}, 113014 (2012),
  doi:10.1088/1367-2630/14/11/113014.

\bibitem{Amoretti:2013nv} 
  A.~Amoretti, A.~Blasi, G.~Caruso, N.~Maggiore and N.~Magnoli,
  Eur.\ Phys.\ J.\ C {\bf 73}, no. 6, 2461 (2013),
  doi:10.1140/epjc/s10052-013-2461-3.

\bibitem{Hasan:2010xy} 
  M.~Z.~Hasan and C.~L.~Kane,
  Rev.\ Mod.\ Phys.\  {\bf 82}, 3045 (2010)
  doi:10.1103/RevModPhys.82.3045.

\bibitem{Cho:2010rk} 
  G.~Y.~Cho and J.~E.~Moore,
  Annals Phys.\  {\bf 326}, 1515 (2011)
  doi:10.1016/j.aop.2010.12.011.

\bibitem{Maldacena:1997re} 
  J.~M.~Maldacena,
  Int.\ J.\ Theor.\ Phys.\  {\bf 38}, 1113 (1999)
  [Adv.\ Theor.\ Math.\ Phys.\  {\bf 2}, 231 (1998)]
  doi:10.1023/A:1026654312961, 10.4310/ATMP.1998.v2.n2.a1.

\bibitem{Hartnoll:2009sz} 
  S.~A.~Hartnoll,
  Class.\ Quant.\ Grav.\  {\bf 26}, 224002 (2009)
  doi:10.1088/0264-9381/26/22/224002.

\bibitem{McGreevy:2009xe} 
  J.~McGreevy,
  Adv.\ High Energy Phys.\  {\bf 2010}, 723105 (2010)
  doi:10.1155/2010/723105.

\bibitem{Amoretti:2017xto}
  A.~Amoretti, A.~Braggio, N.~Maggiore and N.~Magnoli,
  Adv.\ Phys.\ X {\bf 2} (2017) no.2,  409.
  doi:10.1080/23746149.2017.1300509.

\bibitem{2013ass}
J.~McGreevy, 
{\it Holography with and without gravity},
Lectures held at the ``2013 Arnold Sommerfeld School on Gauge-gravity duality and condensed matter physics'',
\url{http://www.asc.physik.lmu.de/activities/schools/archiv/2013_asc_school/videos_ads_cmt/mcgreevy/index.html}.

\bibitem{Amoretti:2014kba} 
  A.~Amoretti, A.~Braggio, G.~Caruso, N.~Maggiore and N.~Magnoli,
  JHEP {\bf 1404}, 142 (2014),
  doi:10.1007/JHEP04(2014)142.

\bibitem{Amoretti:2014iza} 
  A.~Amoretti, A.~Braggio, G.~Caruso, N.~Maggiore and N.~Magnoli,
  Phys.\ Rev.\ D {\bf 90}, no. 12, 125006 (2014),
  doi:10.1103/PhysRevD.90.125006.

\bibitem{Deser:1981wh} 
  S.~Deser, R.~Jackiw and S.~Templeton,
  Annals Phys.\  {\bf 140}, 372 (1982)
  [Annals Phys.\  {\bf 281}, 409 (2000)]
  Erratum: [Annals Phys.\  {\bf 185}, 406 (1988)].
  doi:10.1006/aphy.2000.6013, 10.1016/0003-4916(82)90164-6.

\bibitem{Blasi:2010gw} 
  A.~Blasi, N.~Maggiore, N.~Magnoli and S.~Storace,
  Class.\ Quant.\ Grav.\  {\bf 27}, 165018 (2010),
  doi:10.1088/0264-9381/27/16/165018.

\bibitem{Maggiore:2018bxr} 
  N.~Maggiore,
  Eur.\ Phys.\ J.\ Plus {\bf 133}, no. 7, 281 (2018)
  doi:10.1140/epjp/i2018-12130-y.
\bibitem{Maggiore:1992ki} 
  N.~Maggiore and P.~Provero,
  Helv.\ Phys.\ Acta {\bf 65}, 993 (1992).

\bibitem{Blasi:2011pf} 
  A.~Blasi, A.~Braggio, M.~Carrega, D.~Ferraro, N.~Maggiore and N.~Magnoli,
  New J.\ Phys.\  {\bf 14}, 013060 (2012)
  doi:10.1088/1367-2630/14/1/013060.
\bibitem{Nakanishi:1966zz} 
  N.~Nakanishi,
  Prog.\ Theor.\ Phys.\  {\bf 35}, 1111 (1966).
  doi:10.1143/PTP.35.1111.

\bibitem{Lautrup:1967zz} 
  B.~Lautrup,
  Kong.\ Dan.\ Vid.\ Sel.\ Mat.\ Fys.\ Med.\  {\bf 35}, no. 11 (1967).

\bibitem{Asorey:2006pr} 
  M.~Asorey, D.~Garcia-Alvarez and J.~M.~Munoz-Castaneda,
  J.\ Phys.\ A {\bf 39}, 6127 (2006)
  doi:10.1088/0305-4470/39/21/S03.
\bibitem{Karabali:2015epa} 
  D.~Karabali and V.~P.~Nair,
  Phys.\ Rev.\ D {\bf 92}, no. 12, 125003 (2015)
  doi:10.1103/PhysRevD.92.125003.


\bibitem{Hinterbichler:2011tt} 
  K.~Hinterbichler,
  Rev.\ Mod.\ Phys.\  {\bf 84}, 671 (2012)
  doi:10.1103/RevModPhys.84.671.

\bibitem{Rubakov:2004eb} 
  V.~A.~Rubakov,
  ``Lorentz-violating graviton masses: Getting around ghosts, low strong coupling scale and VDVZ discontinuity,''
  hep-th/0407104.

\bibitem{Libanov:2005vu} 
  M.~V.~Libanov and V.~A.~Rubakov,
  JHEP {\bf 0508}, 001 (2005),
  doi:10.1088/1126-6708/2005/08/001.

\bibitem{Fierz:1939ix} 
  M.~Fierz and W.~Pauli,
  Proc.\ Roy.\ Soc.\ Lond.\ A {\bf 173}, 211 (1939),
  doi:10.1098/rspa.1939.0140.

\bibitem{Blasi:2015lrg} 
  A.~Blasi and N.~Maggiore,
  Class.\ Quant.\ Grav.\  {\bf 34}, no. 1, 015005 (2017),
  doi:10.1088/1361-6382/34/1/015005.

\bibitem{Blasi:2017pkk} 
  A.~Blasi and N.~Maggiore,
  Eur.\ Phys.\ J.\ C {\bf 77}, no. 9, 614 (2017),
  doi:10.1140/epjc/s10052-017-5205-y.

\bibitem{Bassetto:1991ue} 
  A.~Bassetto, G.~Nardelli and R.~Soldati,
  {\it ``Yang-Mills theories in algebraic noncovariant gauges: Canonical quantization and renormalization''},
  Singapore, Singapore: World Scientific (1991) 227 p
\bibitem{Fradkin:1997df} 
  E.~S.~Fradkin and M.~Y.~Palchik,
  Phys.\ Rept.\  {\bf 300}, 1 (1998).
  doi:10.1016/S0370-1573(97)00085-9


\bibitem{Amoretti:2013xya} 
  A.~Amoretti, A.~Braggio, G.~Caruso, N.~Maggiore and N.~Magnoli,
  Adv.\ High Energy Phys.\  {\bf 2014}, 635286 (2014),
  doi:10.1155/2014/635286.

\bibitem{Aratyn:1984jz} 
  H.~Aratyn,
  Phys.\ Rev.\ D {\bf 28}, 2016 (1983),
  doi:10.1103/PhysRevD.28.2016.

\bibitem{Aratyn:1983bg} 
  H.~Aratyn,
  Nucl.\ Phys.\ B {\bf 227}, 172 (1983),
  doi:10.1016/0550-3213(83)90148-7.



\end{thebibliography}
\end{document}